\documentclass[3p,times]{elsarticle}

\usepackage{ecrc}
\usepackage{graphicx}
\newcommand{\pp}{$p$-$p$}
\newcommand{\pbpb}{$Pb$-$Pb$}
\def\ttt#1{\texttt{\small #1}}

\volume{00}

\firstpage{1}

\journalname{Procedia Computer Science}

\runauth{}

\jid{procs}

\jnltitlelogo{Procedia Computer Science}

\usepackage{amssymb}

\usepackage[figuresright]{rotating}

\begin{document}

\begin{frontmatter}



\dochead{}

\title{ALICE potential for direct photon measurements in \pp\ and \pbpb\ collisions }


\author{Rapha\"elle Ichou, for the ALICE collaboration}

\address{Subatech, 4 rue Alfred Kastler, BP 20722 Nantes Cedex 3, France}

\begin{abstract}
The production of direct photons, not coming from hadron decays, at large transverse momentum $p_T>2$~GeV/c in proton-proton collisions at the LHC, is an interesting process to test the predictions of
perturbative Quantum Chromodynamics at the highest energies ever and to put constraints on the gluon density in the proton. Furthermore, they provide a baseline reference for quark-gluon-plasma studies in \pbpb\ collisions. We will present the experimental capabilities of the ALICE electromagnetic calorimeter EMCal to reconstruct the direct and isolated photon spectra in \pp\ and \pbpb\ collisions.
\end{abstract}


\end{frontmatter}



\section{Physics motivations}
\label{pm}

The study of photon production at large transverse momenta ($p_T \gg \Lambda_{QCD}$ = 0.2 GeV) in hadronic interactions is a valuable testing ground of the perturbative regime of Quantum Chromodynamics (pQCD)~\cite{Aurenche:2006vj}. At the LHC, photons will allow one to confront the data with pQCD predictions at energies never reached before. Since these photons come directly from parton-parton hard scatterings, they allow one to constrain the gluon distribution function in the proton at small parton momentum fraction $x$ = $p_{parton}$/$p_{proton}$~\cite{Ichou:2010wc}. Also, photons produced in \pp\ collisions provide a ``vacuum" baseline reference for the study of their production rates in nucleus-nucleus collisions. 
In \pbpb\ collisions, the measurement of direct photons allows one to search for thermal photons emission at 1-5~GeV/c and to study jet-quenching in back-to-back $\gamma$-jet correlations.




The measurement of direct photon production is complicated by a large $\gamma$ background from hadrons, specially from $\pi^0$ mesons, which decay into two photons.
At high $p_T$ in \pp\ collisions, there are between 10 and 100 (at 100~GeV/c and 10~GeV/c, respectively) times more $\pi^0$ than direct $\gamma$ and from 2 to 10 times more (depending on the amount of jet quenching) in \pbpb\ collisions. 
Furthermore, above 10~GeV/c, the two $\pi^0$ decay photons start to merge into a single cluster in the EMCal~\cite{ppremcal} and thus cannot be identified as a $\pi^0$ via $\gamma$-$\gamma$ invariant mass analysis.
Above 10~GeV/c, the identification method is based on the shape of the shower, which allows to identify photons up to $\sim$45~GeV/c in the EMCal.
At higher $p_T$, as the two $\gamma$ totally overlap, one requires the photons to be isolated from any hadronic activity within a given distance around its direction, to remove the decay photons
produced inside jets. The corresponding measurements are then dubbed \textit{isolated} photons. 
~\\
In ALICE, photons can be measured by two electromagnetic calorimeters, EMCal, a Pb-Sc calorimeter, located at mid-rapidity at -0.7~$<~\eta~<$~0.7, which will have a total acceptance of $\Delta \phi =$~107$^{0}$ in azimuth and PHOS (PHOton Spectrometer), a lead tungstate crystal calorimeter, located at mid-rapidity at -0.12~$<~\eta~<$~0.12 and covering $\Delta \phi =$~100$^{0}$.
ALICE can also measure photons from their conversion in electron-positron pairs inside the central tracker.
~\\

There are 4 general strategies to measure direct photons: 
\begin{enumerate}
\item{A statistical method, which consists in subtracting from the total inclusive photon spectrum the estimated decay-photon contributions of the measured $\pi^0$ and $\eta$ spectra.}
\item{A tagging method, which consists in removing pairs with mass compatible with $\pi^0$ from the total photon candidate spectrum and correcting for true photon losses.}
\item{Via conversions ($\gamma \to e^+e^-$) in the material within the central tracking system, before the calorimeter.}
\item{Via event-by-event isolation cuts (and statistical subtraction of expected background due to isolated $\pi^0$).}
\end{enumerate}

In this work, we will focus on method 4. A standard isolation requirement, in \pp\ collisions, is that within a cone around the $\gamma$ direction defined in pseudo-rapidity $\eta$ and azimuthal angle $\phi$ by $R = \sqrt{(\eta - \eta_{\gamma})^2 + (\phi - \phi_{\gamma})^2}$, the accompanying hadronic transverse energy is less than a fixed fraction $\varepsilon$ (e.g. often 10\%) of the photon's $p_T$. $R$ is usually taken between 0.4 and 0.7. 
In \pbpb\ collisions, due to large underlying-event background, we use a simple fixed cut on $p_T$.
Isolation enables one to reject a large fraction of the $\pi^0$ decay photons, and access to true isolated photons, prompt photons, produced directly in a hard process via :
(i) quark-gluon Compton scattering $q g \rightarrow \gamma q$, (ii) quark-antiquark annihilation $q \bar q \rightarrow \gamma g$.

\section{Isolated photons measurement in \pp\ collisions with EMCal}
\label{isopp}

\begin{figure}[htbp!]
  \centering
    \includegraphics[height=7.7cm, width=7.2cm, clip=true]{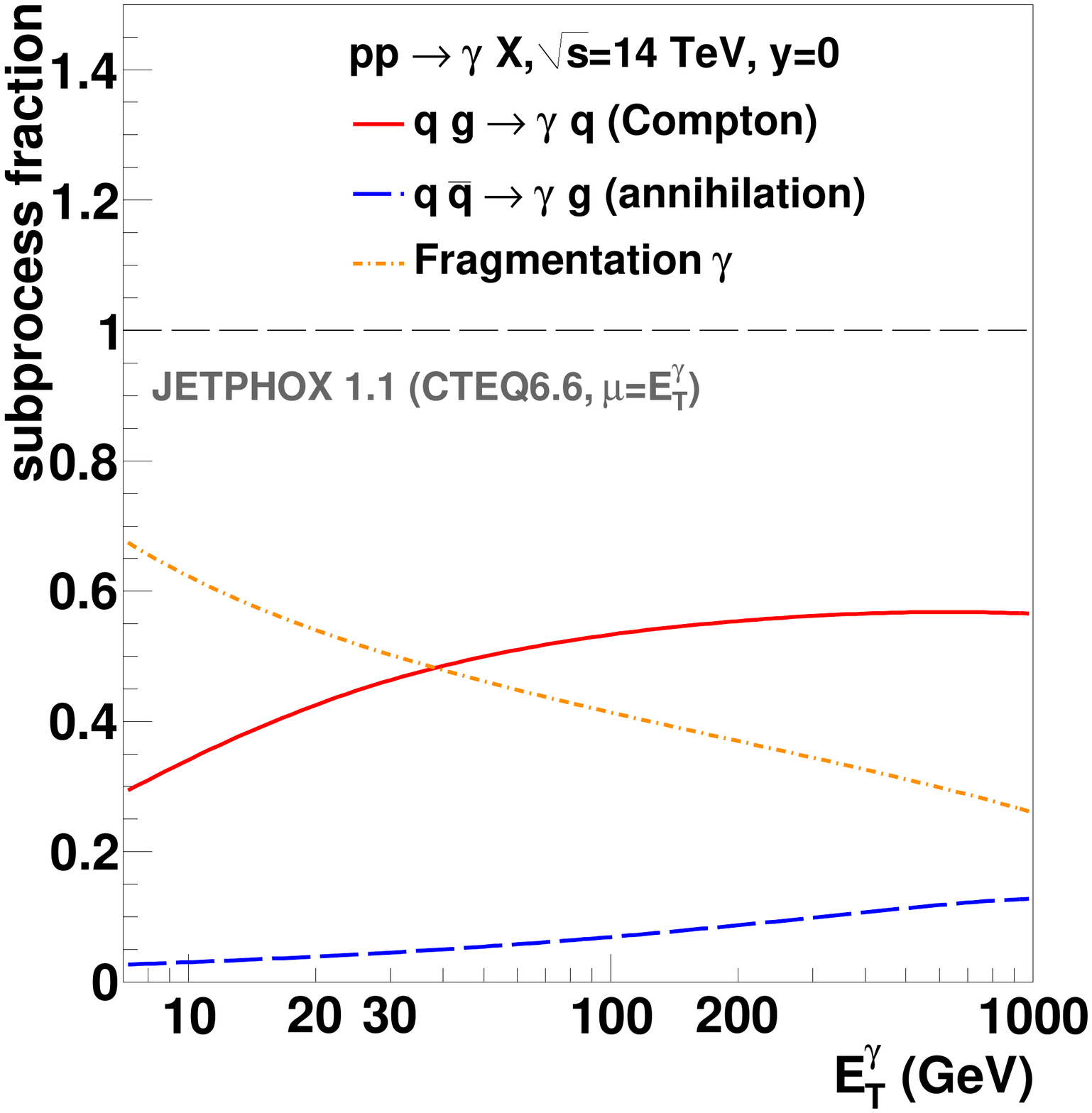}
    \includegraphics[height=7.7cm, width=7.2cm, clip=true]{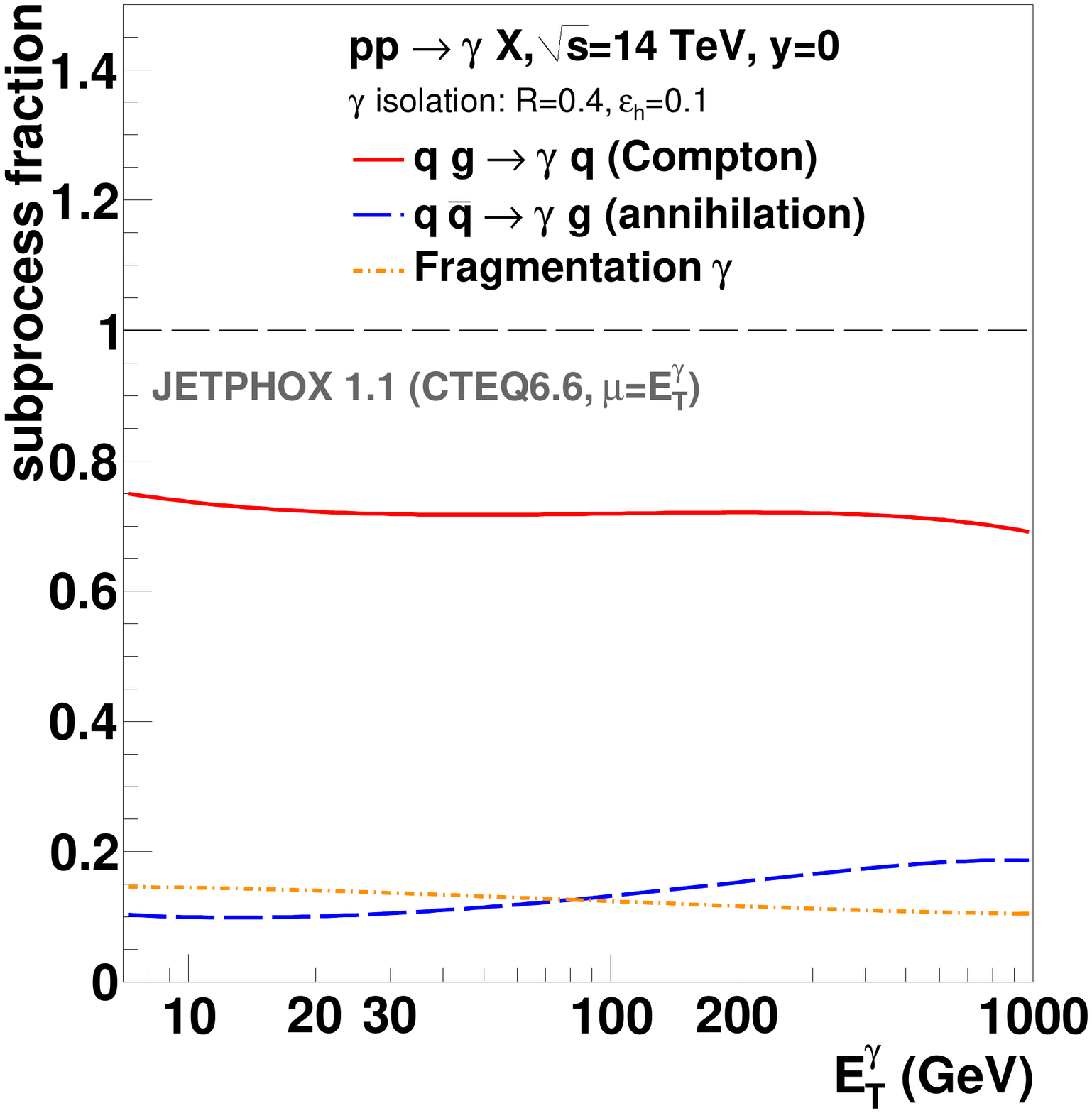}
          \caption{Relative contributions of the quark-gluon Compton, $q\bar{q}$ annihilation and fragmentation subprocesses in NLO direct photon production at LHC midrapidity, obtained with \ttt{JETPHOX} (\ttt{CTEQ6.6} PDF, $\mu$~=~$E_T$, and \ttt{BFG-II} FFs), in the inclusive case (left) and isolated case (right).}
      \label{subproc}
\end{figure} 
The \ttt{JETPHOX}~\cite{jetphox} program (for JET-PHOton/hadron X-sections) allows one to calculate perturbatively the $\gamma$ cross-sections at Next to Leading Order accuracy. It includes both the fragmentation and the Compton and annihilation components and isolation cuts can be applied at the parton level.
Figure~\ref{subproc} left shows the relative contributions of each one of the three direct photon subprocesses in \pp\ (Compton, annihilation and fragmentation) to direct photon production at LHC (y = 0) as a function of the photon $E_T$. They have been obtained setting all scales to $\mu~=~E_T$, and using the \ttt{CTEQ6.6} parton densities and the \ttt{BFG-II} parton-to-photon FFs for the fragmentation photons. The Compton process is dominant above 45~GeV/c, whereas for lower $E_T$, fragmentation photons dominate the spectrum.
Figure~\ref{subproc} right shows the subprocesses contributions to the isolated photon cross section. At variance with the inclusive case, we can see that a very significant part of the fragmentation component is suppressed after applying typical isolation cuts (R = 0.4, $\varepsilon_h$~=~0.1). The Compton process now clearly dominates the photon yield and accounts for about 3/4 of the isolated production for all $E_T$'s.
Although isolation cuts reduce the fragmentation contribution, a fraction of fragmentation photons with z~$\geq$~1/(1~+~$\varepsilon_h$) survive the cuts. A typical isolation energy cut of $\varepsilon_h$~=~0.1, corresponding to 1/(1 + $\varepsilon_h$)~$>$~0.9, suppresses about 60-80\% of fragmentation photons. 

As isolated fragmentation photons, the isolated $\pi^0$ fraction is represented by high-$z$ $\pi^0$ which carry a large fraction of the jet energy ($z$~=~$p_{hadron}/p_{parton}$) and are thus isolated from accompanying hadronic activity. High-$z$ isolated $\pi^0$ are thus considered as the main background in the isolated photons measurement.
\begin{figure}[htbp!]
    \centering
    \includegraphics[height=7.2cm, width=7cm, clip=true]{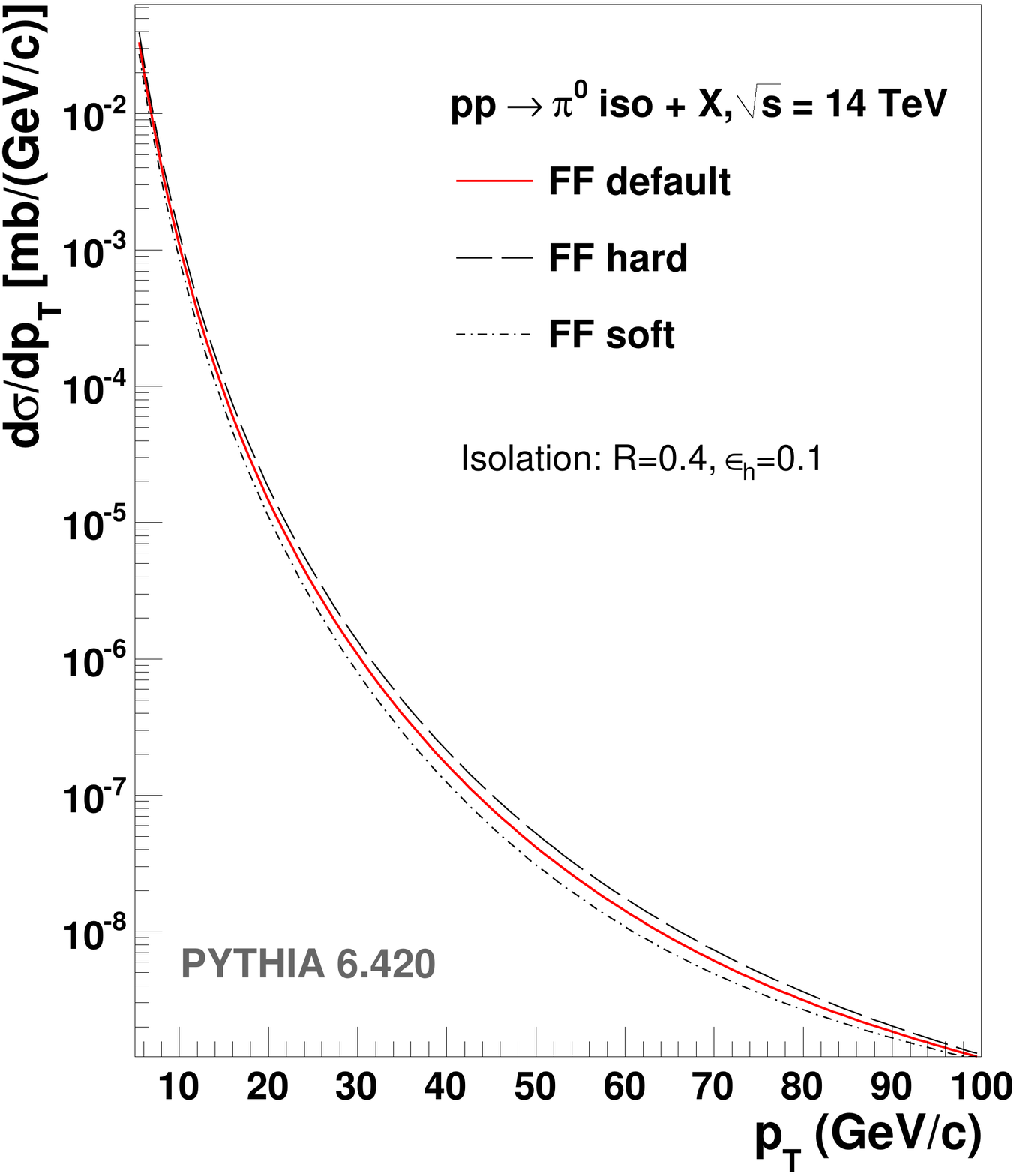}
    \includegraphics[height=7.2cm, width=7cm, clip=true]{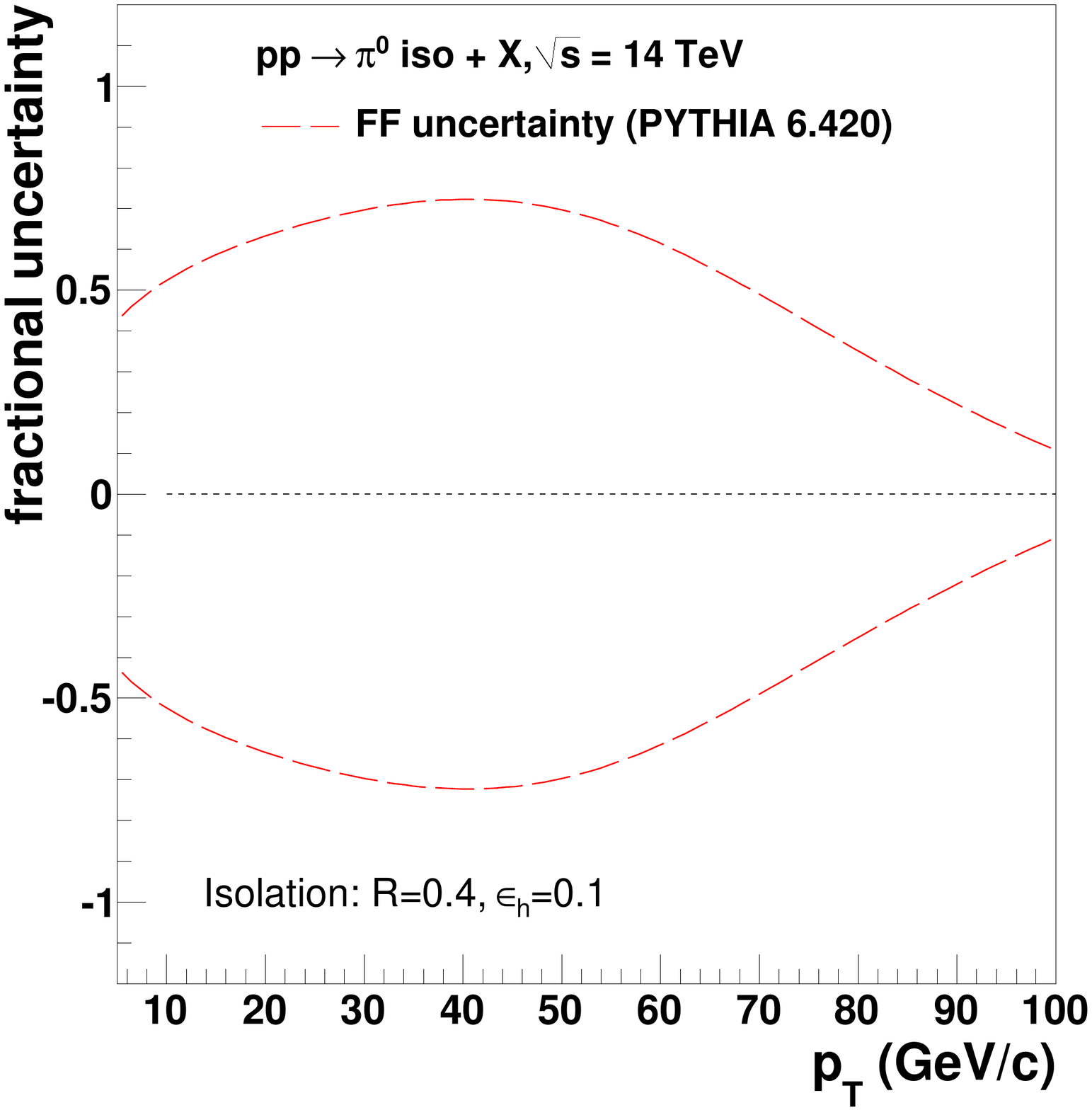}
          \caption{Left: Isolated $\pi^0$ $p_T$ spectrum at midrapidity, obtained with two tunes of $\pi^0$ FFs in \ttt{PYTHIA} (dashed lines, the solid curve is their average) in \pp\ collisions at $\sqrt{s}$~=~14~TeV. Right: Fractional differences between the isolated $\pi^0$ spectrum for each FF tune and their average.}
      \label{pi0}
\end{figure} 
Figure~\ref{pi0} left shows the isolated $\pi^0$ cross-section ($R$~=~0.4, $\varepsilon$~=~0.1) obtained for two different \ttt{PYTHIA}~\cite{Sjostrand:2006za} tunes : Perugia \textit{hard} and \textit{soft}~\cite{Skands:2010ak}, which have
different hardness for the parton-to-$\pi^0$ fragmentation. The expected isolated $\pi^0$ spectrum has systematic uncertainties of up to 70\%, depending on the $\pi^0$ FF used (Figure 2 right).
\begin{figure}[htbp!]
   \centering
    \includegraphics[height=7.3cm, width=7cm, clip=true]{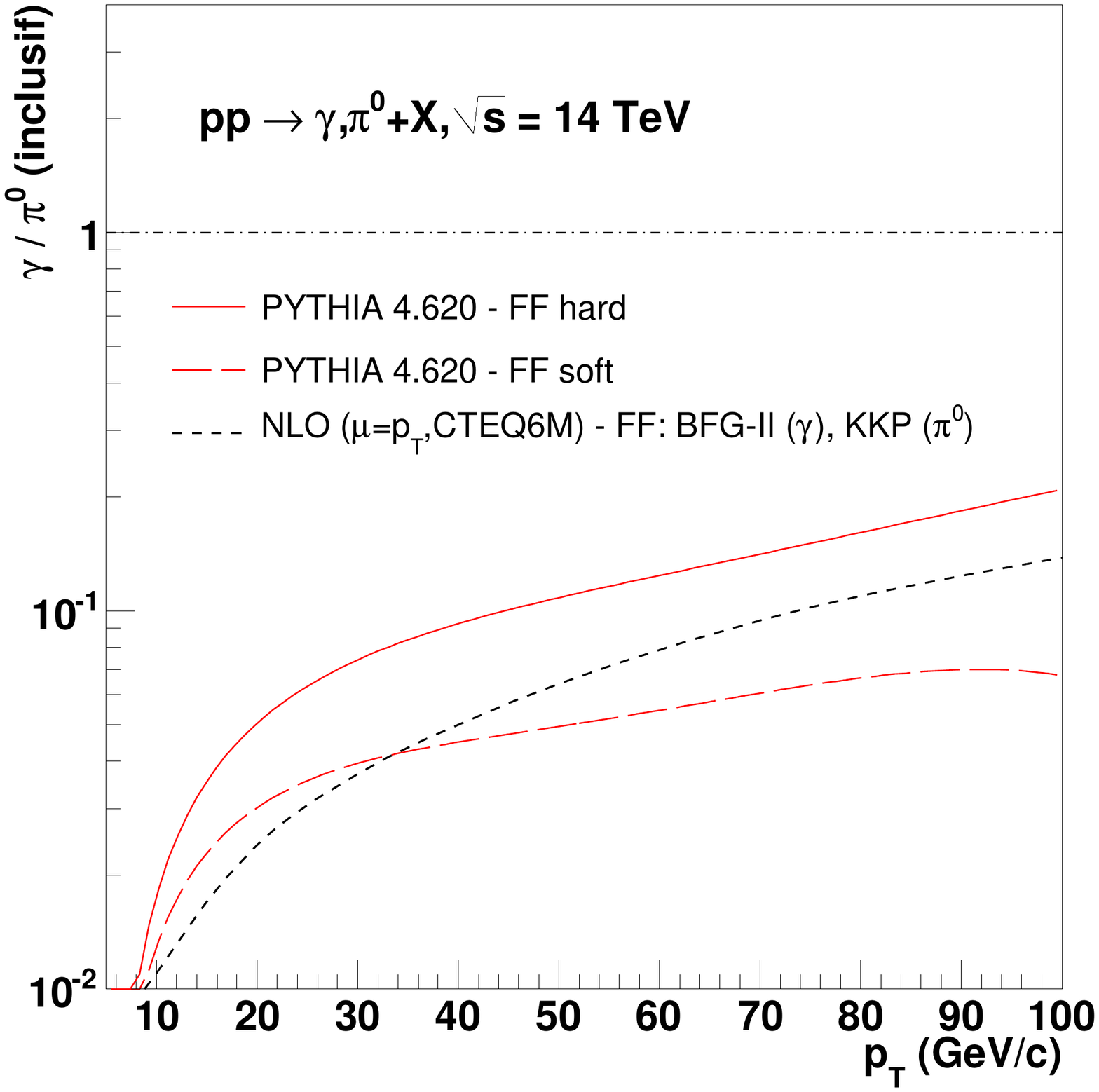}
    \includegraphics[height=7.3cm, width=7cm, clip=true]{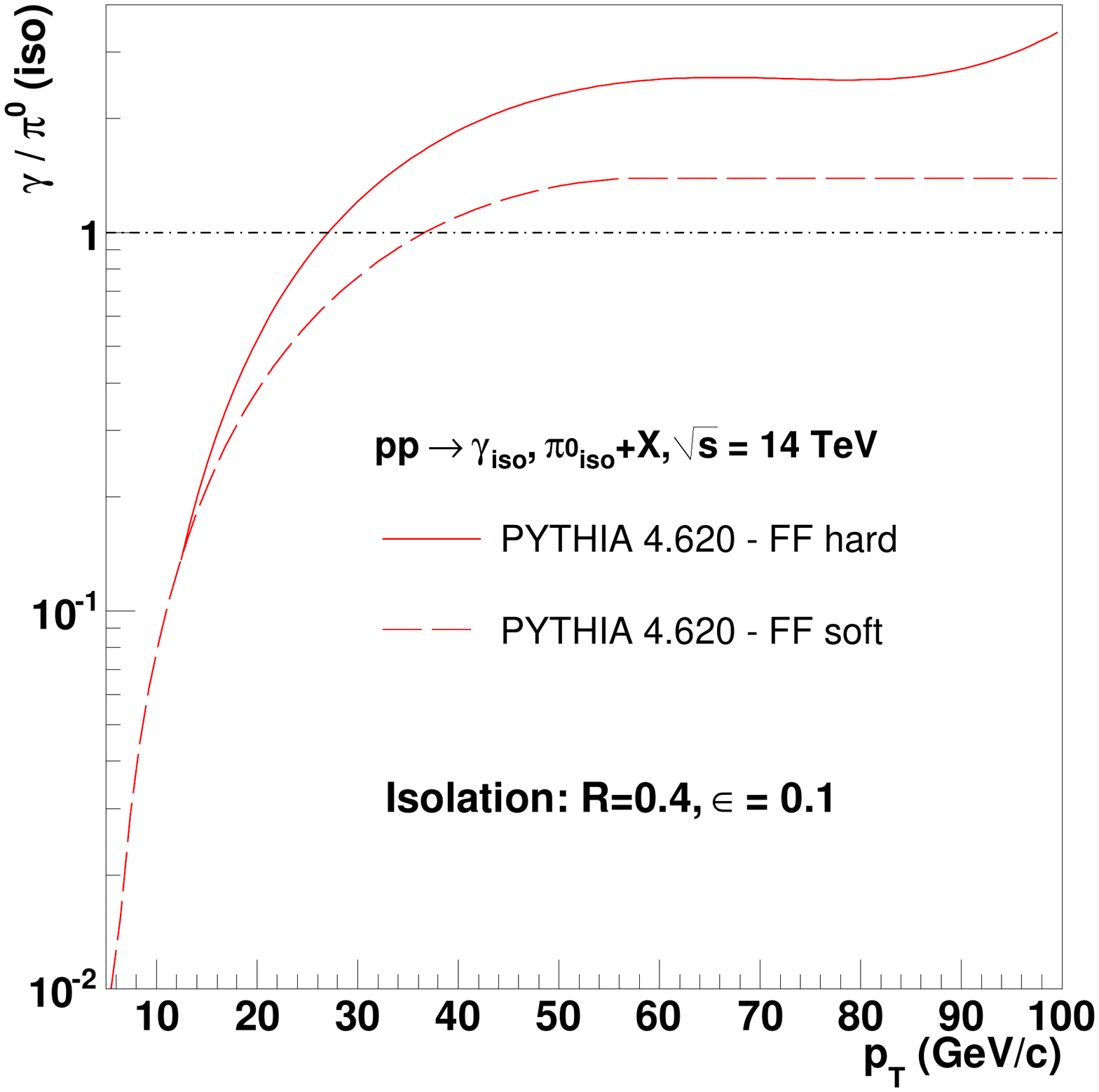}
          \caption{Signal over background (S/B) in the inclusive case (left) and in the isolated ($R$~=0.4, $\varepsilon$=0.1) case (right), for both \textit{hard} and \textit{soft} tunes, as a function of $p_T$, for \pp\ collisions at $\sqrt{s}$~=~14~TeV, at mid-rapidity.}
      \label{setb}
      \end{figure} 

~\\

After obtaining the isolated photon and $\pi^0$ cross-section with \ttt{PYTHIA}, the signal over background (S/B) ratio is now studied. Figure~\ref{setb} left shows the S/B in the inclusive case, and right the S/B with isolation cuts applied ($R$~=~0.4, $\varepsilon$~=~0.1), for \textit{hard} and \textit{soft} FF tunes, for \pp\ collisions at $\sqrt{s}$ = 14 TeV. 
The S/B stays smaller than 1 over the whole $p_T$ range. After applying isolation cuts, the S/B becomes greater than the unit at around 30~GeV/c.
Isolation allows one to enhance significantly the S/B ratio, up to a factor 20 at 100~GeV/c.

\section{Isolated photons measurement in \pbpb\ collisions with EMCal}
\label{isopbpb}

\begin{figure}[htbp]
   \centering
    \includegraphics[height=8.5cm, width=8cm, clip=true]{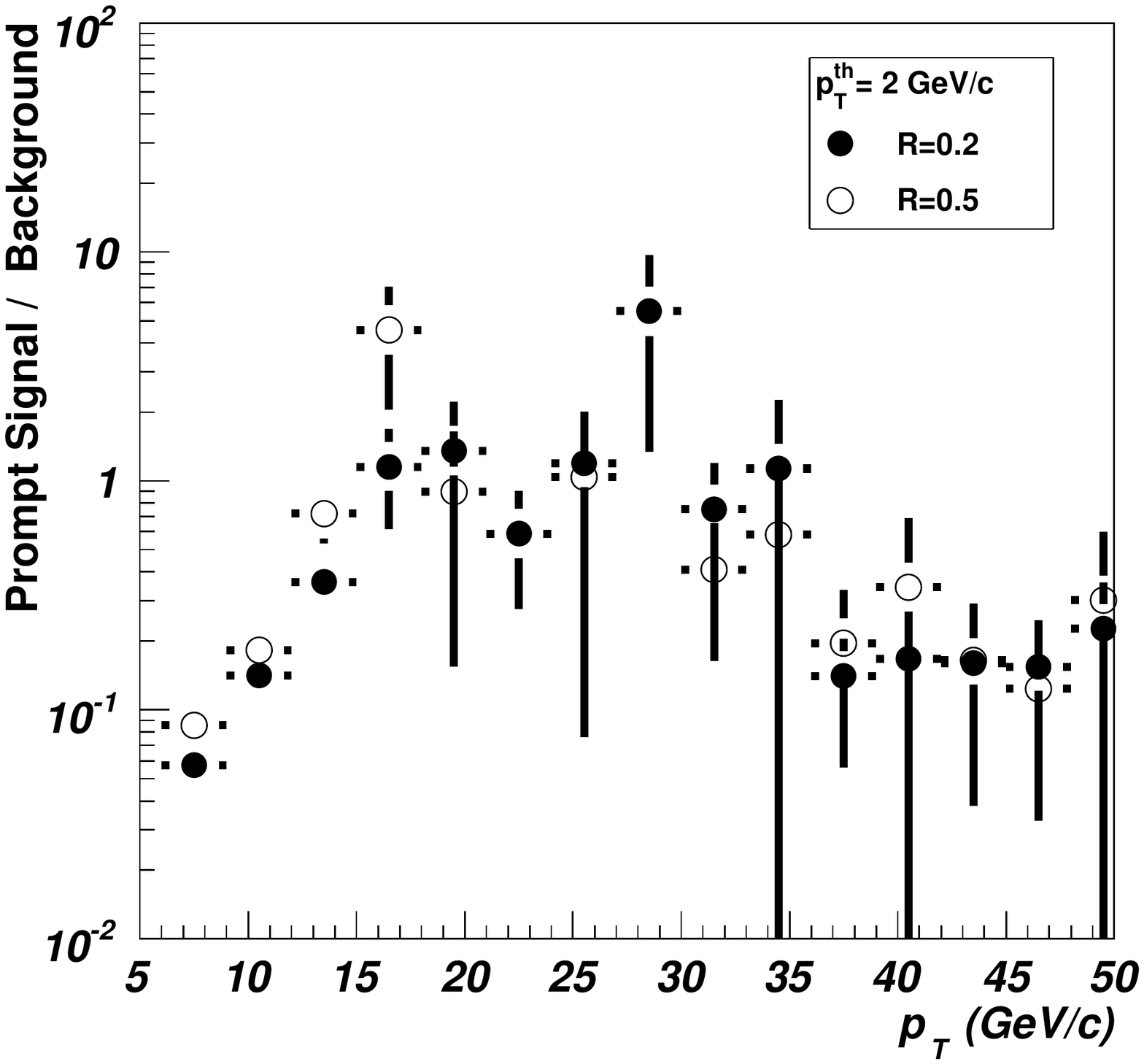}
    \includegraphics[height=8.5cm, width=8cm, clip=true]{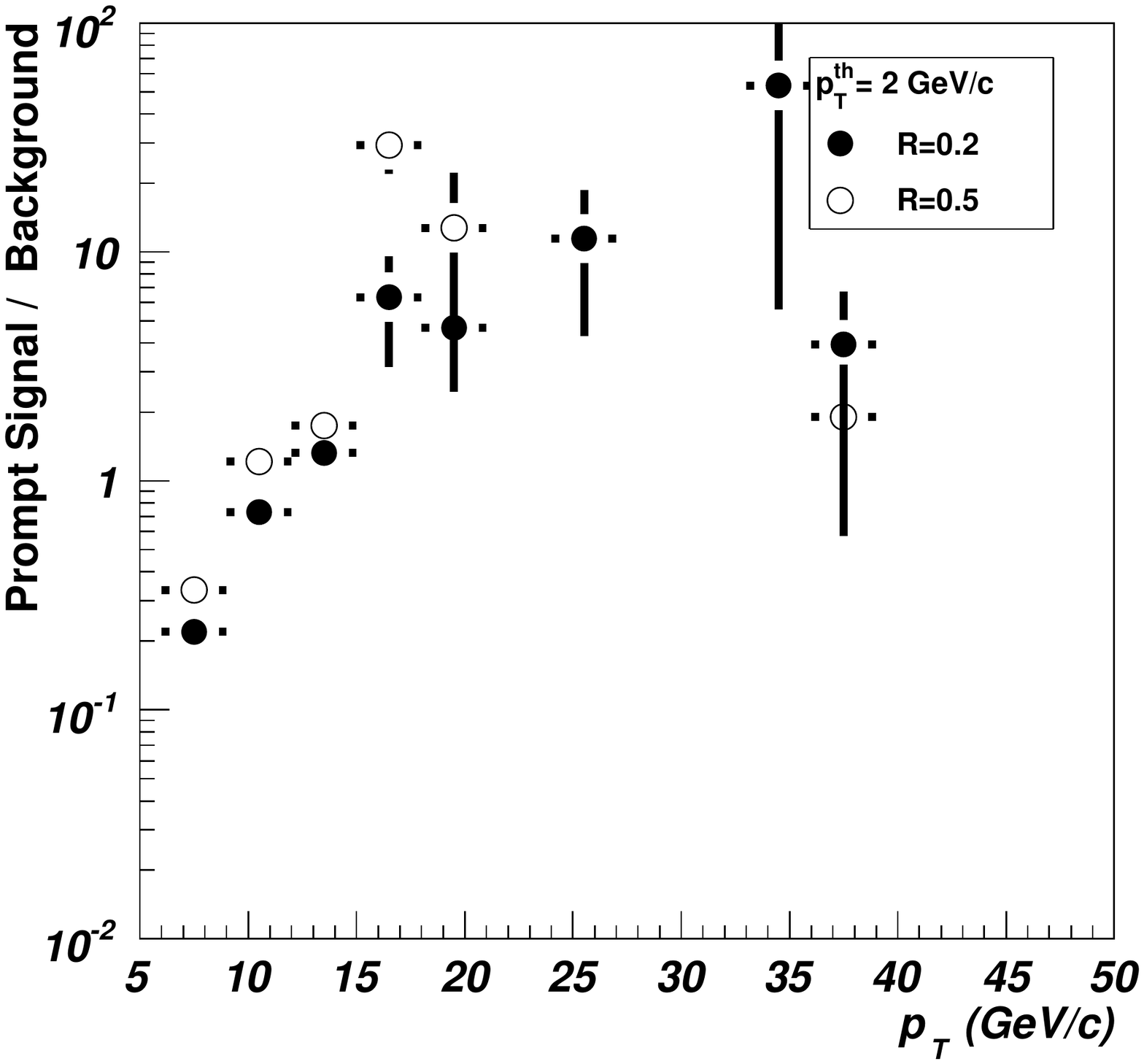}
          \caption{Signal over background (S/B) in the isolated case ($R$~=0.2 and 0.5, ${p_T}_{cut}$~=~2~GeV/c) without (left) and with (right) quenching ($\hat{q}$~=~50~GeV$^2$/fm), as a function of $p_T$, for \pbpb\ collisions at $\sqrt{s}$~=~5.5~TeV, at mid-rapidity~\cite{Conesa:2007gr}.}
      \label{pbpb}
      \end{figure}

For the nominal luminosity in \pbpb\ collisions at 5.5~TeV, ALICE EMCal can reconstruct jets up to 200 GeV/c, $\pi^0$'s up to 50 GeV/c (80 GeV/c) with (without) quenching and direct photons up to 50 GeV/c. 
In Figure~\ref{pbpb}, we show the isolated $\gamma$/$\pi^0$ ratio, computed with \ttt{PYTHIA} $\gamma$-jet (signal) and jet-jet ($\pi^0$ background) events, simulated and reconstructed in full ALICE, and applying isolation cuts ($R$~=~0.2 and $R$~=~0.5, ${p_T}_{cut}$~=~2~GeV/c), for \pbpb\ collisions at 5.5~TeV, with ($\hat{q}$~=~50~GeV$^2$/fm) and without quenching. S/B is significantly enhanced when quenching is applied (Figure~\ref{pbpb} right), becoming bigger than one after 15~GeV/c, and so making the isolated photon measurement in \pbpb\ possible.

\bibliographystyle{elsarticle-num}
\bibliography{<your-bib-database>}



%

%

\end{document}